\documentclass[3p]{elsarticle}

\usepackage{amssymb}
\usepackage{amsmath}
\usepackage{lineno}

\journal{Applied Mathematics and Computation}

\begin{document}


\begin{frontmatter}

\title{Restoring spatial cooperation with myopic agents in a three-strategy social dilemma}

\author[label1]{Hsuan-Wei Lee}
\author[label2]{Colin Cleveland}
\author[label3]{and Attila Szolnoki}
\cortext[mail1]{Correspondence should be addressed to: hwwaynelee@gate.sinica.edu.tw or szolnoki.attila@ek-cer.hu.}

\address[label1]{Institute of Sociology, Academia Sinica, Taiwan}
\address[label2]{Department of Informatics, King's College London, London, UK}
\address[label3]{Institute of Technical Physics and Materials Science, Centre for Energy Research, P.O. Box 49, H-1525 Budapest, Hungary}

\begin{abstract}
Introducing strategy complexity into the basic conflict of cooperation and defection is a natural response to avoid the tragedy of the common state. As an intermediate approach, quasi-cooperators were recently suggested to address the original problem. In this study, we test its vitality in structured populations where players have fixed partners. Naively, the latter condition should support cooperation unambiguously via enhanced network reciprocity. However, the opposite is true because the spatial structure may provide a humbler cooperation level than a well-mixed population. This unexpected behavior can be understood if we consider that at a certain parameter interval the original prisoner's dilemma game is transformed into a snow-drift game. If we replace the original imitating strategy protocol by assuming myopic players, the spatial population becomes a friendly environment for cooperation. This observation is valid in a huge region of parameter space. This study highlights that spatial structure can reveal a new aspect of social dilemmas when strategy complexity is introduced.
\end{abstract}


\end{frontmatter}

\section{Introduction}

If I exploit my supporting partner, then I can enjoy a higher payoff at their expense. Even in the absence of retaliation or revenge, the extra income could only be a short-term success because an intelligent neighbor will adopt my more successful strategy. Hence, we both become empty handed eventually. This is the key element which establishes the so-called network reciprocity, first identified by Nowak and May \cite{nowak_n92b}. This mechanism, which is practically an enhanced form of direct reciprocity \cite{nowak_s06}, has been tested and justified by several forthcoming works \cite{szabo_pre05,feng_s_csf23,liao_hm_amc23,ding_r_csf23,yu_fy_csf22,dercole_srep19} and became a starting element when different consequences of interaction topologies were studied on the evolution of cooperation \cite{floria_pre09,du_cp_amc23,liu_lj_rsif22,kang_hw_pla21}. Notably, there are certain mechanisms, like searching for a more supportive environment \cite{helbing_pnas09,xiao_zl_njp20}, avoiding toxic neighbors via migration \cite{xiao_sl_epjb22,lee_hw_c22}, giving up the concept of uniform supporting and focusing on a specific neighbor \cite{szolnoki_amc20,cao_xb_pa10,he_jl_pla22,lee_hw_pa21}, or applying interaction stochasticity toward different neighbors \cite{chen_xj_pre08b,gao_sp_pla20}, which cannot be interpreted in a well-mixed population. Therefore, assuming a certain structure of interaction is inevitable for them.

An alternative way to improve the general cooperation level could be the introduction of competing strategies whose presence can mitigate the fast success of defection and reveal the mutual benefit of cooperation \cite{szolnoki_pre14b,fu_mj_pa21,quan_j_csf21,szolnoki_njp16,sun_xp_pla23}. It is almost impossible to summarize the various suggestions about strategy complexity that could be useful against defection. Instead, we just refer to topical reviews \cite{perc_jrsi13,perc_pr17} where interested readers can easily find further examples. However, a rather general feature of these approaches is that a new strategy, individual or global, should pay an extra cost for the mentioned goal. This observation is valid whether we apply punishment \cite{szolnoki_pre11b,battu_srep23,flores_jtb21,lee_hw_amc22}, or reward \cite{wang_xj_csf22,fu_mj_pa21,szolnoki_epl10}, individually, or collectively via an institution \cite{li_my_csf22,ohdaira_srep22,gao_sp_pre20,sun_xp_amc23}. An interesting new model was recently introduced by Chen {\it et al.} who applied a so-called quasi-cooperator strategy that lies between unconditional cooperation and defection \cite{chen_q_csf22}. In particular, the new strategy, also called a negative cooperator, exploits unconditional cooperator players less than a defector, and in parallel, it is more protected against the latter strategy. The authors observed that the presence of quasi-cooperators is helpful to block the fast and easy invasion of defectors, hence providing a better condition to maintain cooperation.

Notably, the consequence of quasi-cooperators was only tested in an infinite well-mixed population. This is surprising, because, as we argued earlier, a structured population could offer an extra supporting condition for cooperation \cite{szolnoki_csf22,hua_sj_csf23}. It is also a frequent observation that a mechanism is inefficient in a random system while it has positive consequences in a structured population \cite{xie_k_csf23,szolnoki_srep21,he_jl_pla23}. To fill this gap, in this study, we extend the original model to a spatial version setup and find a rather unexpected result. More precisely, we obtain that a spatial version yields worse results than in a random population; hence fixed partners are detrimental to the evolution of cooperation now.  

The explanation of this surprising system behavior is based on the fact that the general parametrization of the payoff matrix allows a transformation of the original prisoner's dilemma situation to a case when a snow-drift game characterizes the relation of certain competing strategies. However, this social dilemma game can block the spatial clustering of cooperators if the microscopic dynamics are executed via imitation of the more successful strategy protocol \cite{hauert_n04}. Nevertheless, if we apply alternative microscopic dynamics, where players explore the potential alternatives independently of the neighboring strategies, then stable partners can become a cooperation supporting condition again. This observation highlights that we should be careful when we try to identify cooperation-supporting conditions in general, because the system behavior may depend on the applied microscopic dynamics.

The remainder of this paper is organized as follows. In the next section we define the three-strategy model. We specify the applied microscopic updating protocols, which are imitation of a neighboring strategy based on pairwise comparison \cite{szabo_pre98} and the extension of so-called myopic agents where players try to maximize their own payoff upon a strategy change while assuming that the neighboring strategies remain unchanged \cite{sysiaho_epjb05}. In Section~\ref{result} we compare the fractions of strategies obtained by assuming well-mixed and structured populations. Furthermore, different strategy updating protocols are also tested. To achieve a robust and generally valid observation, we extend the parameter ranges to all possible values. Finally, we summarize the main conclusions and discuss their potential implications in Section~\ref{conclusion}.

\section{Model}

Our system contains three available competing strategies. In addition to unconditional cooperators ($C$) and defectors ($D$), we assume a third strategy that can be considered as an intermediate strategy between the traditional extreme states \cite{chen_q_csf22}. These quasi-cooperators ($QC$) are not as devoted to collective success as cooperators, but they still make some effort. As a result, they still benefit more than the partner when faced with a pure cooperator, but the latter is exploited less than from an interaction with a pure defector. Furthermore, a $QC$ player is less sensitive to the exploitation of a defector, who in turn can collect a reduced income from the interaction with this third-type player. The specific elements of the payoff matrix are the following	
\begin{center}
	\begin{tabular}{r|c c c}
		& $C$ & $QC$ & $D$\\
		\hline
		$C$\,& $1$ & $1-\alpha$ & $-\delta$\\
		$QC$ & $1+\alpha$ & $\rho$ & $m-\delta$\\
		$D$\,& $b$ & $b-\lambda$ & $0$\\
	\end{tabular}\,\,.
\end{center}
\vspace{4mm}
Here parameter $b$ represents the traditional $T$ temptation to defect while the negative value of $\delta$ determines the $S$ sucker's payoff. Typically, the $R$ reward for mutual cooperation is chosen to be 1, while the $P$ punishment for mutual defection is 0. In this way, the $T>R>P>S$ rank ensures that the interaction of $C$ and $D$ players is always described by the well-known prisoner's dilemma game. The additional parameters that characterize the income of the new strategy include the following. Parameter $\alpha$ reflects the ``moderate collective effort'' of $QC$ player, which implies an extra income against a pure cooperator whose income is reduced by the same quantity. Evidently, smaller mutual effort is less fruitful than pure cooperation, which is represented by a $\rho<1$ reward between a $QC$ pair. Regarding the relation of $QC$ and $D$ players, the latter strategy cannot exploit the former as efficiently as a pure cooperator; hence the temptation value is reduced by a $\lambda$ value. Contrastingly, $QC$ is capable to gain a larger $S$ value, which is increased by $m$ compared to the pure $C$ sucker's payoff. In agreement with a previous study by Chen {\it et al.}, we fix the parameters $\rho=0.5$ and $\delta=0.1$ while varying the values of the remaining four parameters in full scale.

As already noted, the relation between $C$ and $D$ strategies always fulfills the criterion of the prisoner's dilemma game. However, the fundamental relation of $QC$ and $D$ strategies is not as straightforward, and it may depend on the values of control parameters. For example, if $\lambda$ is small, the relation of payoff values establishes the $T>R>S>P$ rank, which recalls the traditional snow-drift game \cite{li_k_csf21}. As we demonstrate later, this change is significant to how a spatially structured population behaves.
 
The above-defined model was already studied in an infinite well-mixed population using a replicator equation~\cite{chen_q_csf22}. The key observation was that the presence of quasi-cooperators can help the system avoid the tragedy of the common state because they can weaken the defector's power, thus indirectly promoting the evolution of cooperation.

In this study, we extend this model by assuming spatial populations. Specifically, we assume that players are staged on a $L \times L$ square lattice where periodic boundary conditions are applied. In agreement with previous works~\cite{szabo_pre05,szabo_pre98}, we assume that players are motivated to reach a more favorable individual payoff by imitating a more successful neighbor. Further, we assume that the microscopic dynamics follow the so-called strategy imitation based on pairwise-comparison. If a randomly chosen player $i$, having strategy $s_i$, selects a neighbor $j$, who represents strategy $s_j$, then the player $i$ adopts strategy $s_j$ with a probability
\begin{equation}
	\Gamma(s_j \to s_i)=\frac{1}{1+\exp[(\Pi_{i}-\Pi_{j}) /K]}\,\,.
	\label{fermi}
\end{equation}
Here $\Pi_i$ and $\Pi_j$ denote the payoff values of players $i$ and $j$ collected from the games played with their nearest neighbors, and $K>0$ is the parameter that characterizes noise level. The lower the value of $K$, the more deterministic the adoption of the more successful strategy. In this study we have chosen a representative $K=0.5$ noise level which makes it possible to compare our results with previous observations. If the above-described elementary step is executed $N=L \times L$ times then each player can update her strategy once on average, which establishes a full Monte Carlo (MC) step. During our simulation we used $L=200 - 400$ linear size to gain reliable data where typical running time was 10,000 MC steps.
 
Additionally, we tested an alternative strategy updating rules where the focal player modifies her actual strategy independent of the strategies of the neighbors. Importantly, the focal player assumes that the environment does not change during her trial. In particular, a player $i$ changes her strategy $s_i$ to $s'_i$ with probability
\begin{equation}
	\Gamma(s'_i \to s_i)=\frac{1}{1+\exp[(\Pi_{s_i}-\Pi_{s'_i}) /K]}\,\,,
	\label{myopic}
\end{equation}
where the payoff values are calculated by assuming that the strategies of the nearest neighbors are unchanged~\cite{szabo_jtb12}. This is the so-called myopic best response rule which was previously used for two-strategy systems~\cite{roca_epjb09,li_xy_epjb21,ye_yx_epjb22,shi_j_c21}. In our case a player can try between two other possibilities with equal weights and compare the success with the income originating from the usage of the present strategy. The remaining details of the simulation were identical to those we used for imitation protocol.
 
\section{Results}
\label{result}

First, we show the results obtained at different setups by using the parameter values presented in Ref.~\cite{chen_q_csf22}. In particular, we first measure the fractions of strategies obtained in a system where players interact randomly while imitating a randomly selected partner with a probability defined by Eq.~\ref{fermi}. Second, we use the same imitation protocol, but in a structured population where players are distributed on a square lattice. Third, we use the same structured population, but the microscopic dynamics is the myopic best response protocol defined by Eq.~\ref{myopic}.

\begin{figure}
	\centering
	\includegraphics[width=1\textwidth]{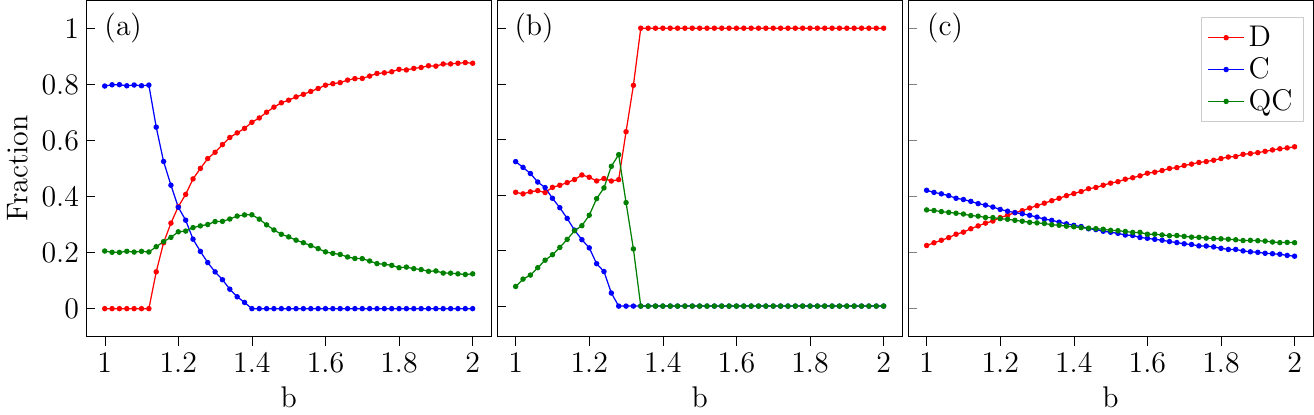}	
	\caption{Fraction of strategies in dependence of temptation obtained at $\lambda=0.7$, $\alpha=0.1$, and $m=0.2$. The left panel shows the result of a well-mixed population where players interact randomly. The middle and right panels show spatial populations using imitation and myopic best response strategy update, respectively. The disadvantage of spatial arrangement for imitation dynamical rule is visible.}
	\label{c2}
\end{figure}

Our first data in Fig.~\ref{c2} shows the stationary fractions of competing strategies in dependence of temptation parameter $b$. It is evident that neither the cooperators nor the quasi-cooperators are viable in a spatial system, and defection performs even better than in a well-mixed case. However, the failure of structured population is restricted only to the applied imitation update protocol. If we employ alternative microscopic dynamics, as shown on the right-hand side panel, then defection is not as vital, even if players are still arranged on a lattice.

The surprising failure of network reciprocity can be explained if we consider the relation of strategies as we vary the values of the payoff parameters. While the relation of $C$ and $D$ strategies still fulfills the criterion of the prisoner's dilemma game, that of $D$ and $QC$ can be different. If $b$ is large enough, their effective payoff values can be described as $T>R>S>P$, which is the case of the snow-drift game. However, this social game can yield interesting system behavior in a spatial population. As noted by Hauert and Doebeli, a spatial structure is not necessarily beneficial for cooperative behavior here~\cite{hauert_n04}. In this case, the compact cooperator clusters cannot evolve, because the $T+S>R+R$ payoff structure prefers a mixed state. The latter can be reached by applying an alternative strategy updating dynamic where the new strategy should not necessarily be that of a neighbor. This criterion is fulfilled when players change their states in an exploratory way rather than through imitation. The myopic best response rule is designed accordingly. Indeed, a significant cooperation level can be reached by using these dynamics~\cite{sysiaho_epjb05,szabo_jtb12}. Our last panel in Fig.~\ref{c2} confirms this expectation because both $C$ and $QC$ strategies survive at the expense of the $D$ strategy in the complete range of the $b$ parameter. 
\begin{figure}[h!]
	\centering
	\includegraphics[width=1\textwidth]{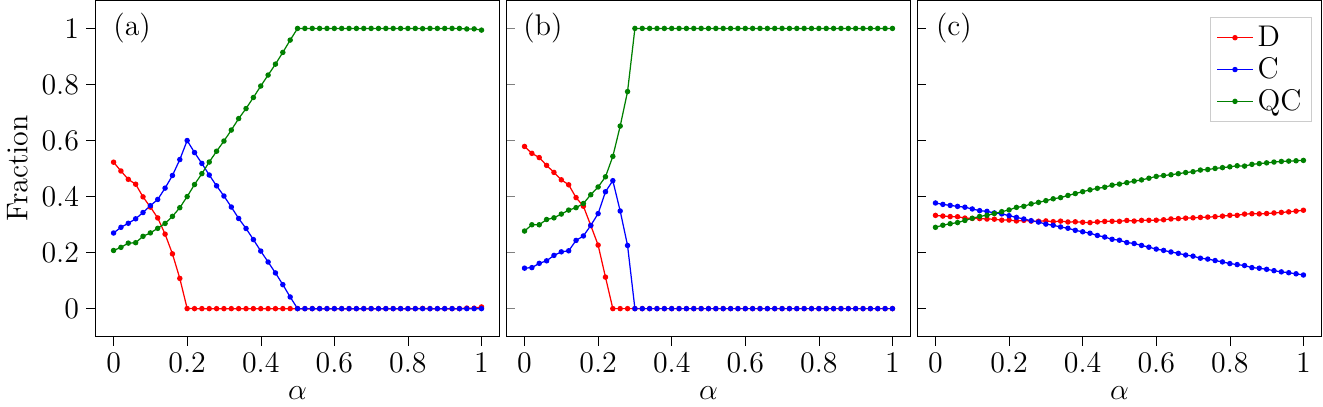}	
	\caption{Fraction of strategies in dependence of $\alpha$ temptation obtained at $\lambda=0.7$, $b=1.2$, and $m=0.2$. The left panel shows the result of a well-mixed population. The middle and right panels show spatial populations using the imitation and myopic best response strategy update, respectively. There is no significant difference between the first two panels because varying $\alpha$ renders the income of $D$ intact, hence the payoff structure does not change.}
	\label{c3}
\end{figure}

The second plot, shown in Fig.~\ref{c3}, presents the strategy fractions according to the $\alpha$ parameter. As illustrated in the first two panels, there is no significant difference between the well-mixed and spatial populations if strategies are updated via imitation. This behavior can be understood because varying $\alpha$ can only change the relation of the $C$ and $QC$ strategies. Therefore, the positive consequence of the presence of quasi-cooperators is not only preserved but is slightly improved for the spatially structured case. As expected, here the usage of myopic best response protocol provides a balanced relation among strategies; hence, defectors can survive even if the value of $\lambda$ is relatively high.

\begin{figure}[h!]
	\centering
	\includegraphics[width=1\textwidth]{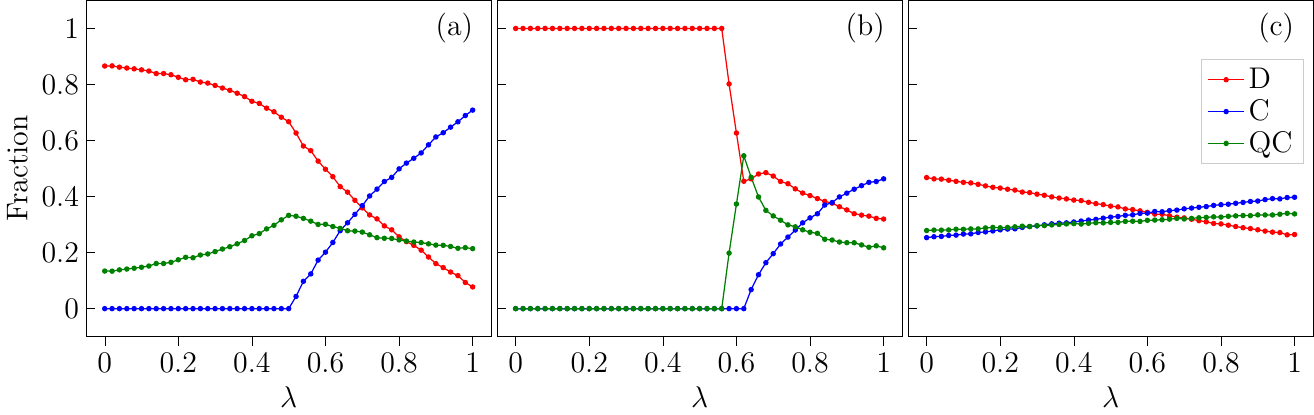}	
	\caption{Fraction of strategies in dependence of $\lambda$ obtained at $b=1.2$, $\alpha=0.1$, and $m=0.2$. Similar to previous plots, the first panel shows the well-mixed case, while the next two panels depict the structured population with imitation and myopic best response protocols. In the low-$\lambda$ interval, the usage of structured interactions is disadvantageous when imitation dynamics is applied. The negative consequence of spatial condition can be fixed by myopic best response dynamics, as shown in the last panel.}
	\label{c4}
\end{figure}

Our third representative plots as shown in Fig.~\ref{c4}, reveal the difference between the well-mixed and structured populations again. A comparison of the first two panels reveals that quasi-cooperators survive in the low-$\lambda$ region in well-mixed population, but defectors prevail in the structured population. In this parameter region, the relation of $D$ and $QC$ fulfills the snow-drift game situation; therefore, the behavioral change observed during the switch from well-mixed to structured situation is consistent with previous findings. As demonstrated in the last panel, the usage of myopic best response can restore the negative consequence of the structured population.

\begin{figure}
	\centering
	\includegraphics[width=1\textwidth]{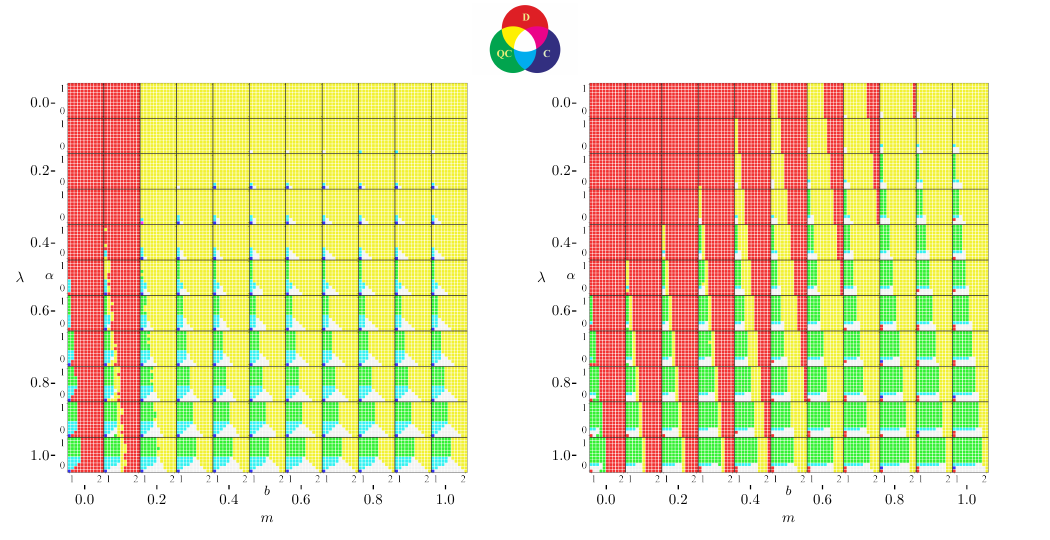}	
	\caption{Stable solutions on the full $\alpha-b$ parameter plane where every mini-panel covers the $0<\alpha<1$, $1<b<2$ intervals. Columns (rows) show different values of $m$ ($\lambda$), as indicated. The color code of different solutions are shown on the top. Accordingly, red indicates the full victory of $D$, green marks the region where $QC$ prevails. Yellow  indicates the parameter space where $D$ and $QC$ strategies are stable. Those regions where all three strategies coexist are marked by white. The left panel shows the well-mixed model while the right depicts the spatial model where strategy imitation drives the evolution.}
	\label{full}
\end{figure}

Evidently, the model contains multiple parameters, including $\alpha$, $b$, $\lambda$, and $m$. Therefore, to demonstrate the robustness of our observations, the complete range of parameter space must be explored. To reveal the characteristic system behavior in this four-dimensional space, we assign specific colors to different solutions where the stationary state contains different strategies. The legend of our color code is shown on the top of Fig.~\ref{full}. Accordingly, the use of red indicates the parameter region where defectors prevail, green indicates when quasi-cooperators dominate the whole system, and blue indicates when simple cooperators win. Further, yellow marks the stable coexistence of $D+QC$, cyan denotes the case when different cooperator strategies survive, and magenta indicates the traditional mixture of $D$ and $C$ players. Finally, white denotes when all three strategies are present in the final state.

To cover the full four-dimensional space, we checked the system behavior at $11^4 = 14641$ combinations of parameters. Evidently, the direct presentation of this four-dimensional space is not possible. Therefore, we display two-dimensional plots on $b-\alpha$ plane and these panels are repeated through different combinations of $m$ and $\lambda$ values. The summary of our results are shown in Fig.~\ref{full} where different columns (rows) depict our results obtained at different values of $m$ ($\lambda$) parameters. The left panel shows the well-mixed population while the right panel demonstrates the spatial system with imitation dynamics. Notably, the myopic best response case is not shown here because all three strategies are present; hence, a similar type of plot would not be meaningful. The difference between the well-mixed and spatial model is striking and confirms our previously reported observations. Namely, the parameter region where pure $D$ strategy prevails is significantly larger when players can only learn from fixed neighbors. Furthermore, this difference is the most spectacular in the low $\lambda$ region where the snow-drift game characterizes the relation of $D$ and $QC$ players.

Interestingly, there is a specific parameter region where spatial imitation can be advantageous. This is the large $\lambda$ - large $m$ corner of the parameter space, where the ``green'' area is significantly larger on the right panel. It indicates that the system behavior presented in Fig.~\ref{c3} is robust and generally valid in a broad range of parameter space.

\section{Conclusions}
\label{conclusion}

Resolving the irreconcilable conflict of cooperation and defection in social dilemmas has been at the frontier of complex system research in the last two decades \cite{quan_j_c19,tao_yw_epl21,zhu_pc_epjb21,fu_mj_pa19,quan_j_pa21}. Defectors who only exploit the competitor cooperators have a clear advantage; however, there are several escape routes for cooperation to survive. One of them is through the use of an alternative strategy that can mitigate the payoff gap. This idea was the motivation for a recently introduced model where an intermediate actor, called a quasi-cooperator, was introduced~\cite{chen_q_csf22}. We revisited the model in this study to test how the competition changes if we assume a structured population.

Importantly, the presence of quasi-cooperators was already helpful in a well-mixed population; therefore, an extended cooperator-supporting mechanism is expected in a spatial system. Several previous model studies have confirmed this belief where network reciprocity is on stage. Interestingly, we observed the opposite system reaction in this study, and the spatial system performed worse than the random condition.

This unexpected behavior can be explained by the subtle relations of strategies. Specifically, the large number of parameters, which characterize the payoff elements of competing strategies, allow the simultaneous emergence of alternative social dilemmas while the basic prisoner's dilemma between the fundamental $C$ and $D$ strategies is still present. Accordingly, the relation between new $QC$ and $D$ strategies may be described by the snowdrift game at certain parameter values. However, this game is of paramount importance, because spatial structure is not necessarily beneficial for cooperative behavior when the snow-drift game payoff parametrization is used. Importantly, this observation is restricted to imitating the more successful strategy update protocol. This is because the reported discrepancy is not valid for other microscopic dynamics, where actors test alternative strategies independently of their neighbors. In short, the spatial system reveals the limited capacity of the imitation rule when network reciprocity cannot work due to the payoff structure of the game. To verify this, we also applied an alternative protocol and the myopic best response dynamics restored the expected system behavior even in a structured population.

Our study highlights the possible danger of payoff matrix complexity when many strategies compete for space. Different types of social dilemmas can best describe the relationship between various competitors. This could be a decisive factor in every model where more than two strategies are present, and their relations are described by different parameters. Notably, this is a different situation from those cases where players simultaneously apply different payoff matrix elements, which could result from varying deceit or environmental factors~\cite{szolnoki_epl14b,shu_lm_prsa22}. 

\vspace{0.5cm}
This research was supported by the National Research, Development and Innovation Office (NKFIH) under Grant No. K142948 and by the Ministry of Science and Technology of the Republic of China (Taiwan) under grant No. 109-2410-H-001-006-MY3.

\bibliographystyle{elsarticle-num-names}

\end{document}